\documentclass[aps,pra,twocolumn,amsmath,amssymb,floatfix,superscriptaddress]{revtex4-1}

\usepackage{graphicx}
\usepackage{physics,bm,braket}
\usepackage{color}
\usepackage{lineno}
\usepackage{bbm}
\usepackage[colorlinks=true,linkcolor=blue,urlcolor=blue,citecolor=blue,anchorcolor=blue]{hyperref}
\usepackage{mathrsfs}
\usepackage{amsmath}
\usepackage{amsfonts}
\usepackage{graphicx,epstopdf}
\usepackage{subfigure}
\usepackage{epsfig}
\usepackage{dcolumn}
\usepackage{bm}
\usepackage{color}
\usepackage{natbib}
\usepackage{xcolor}
\usepackage{braket}
\usepackage{ulem}
\usepackage{float}
\usepackage{lipsum}
\begin{document}

\title{Enhanced response at exceptional points in multi-qubit systems for sensing} 

\author{Tingting Shi}
\affiliation{Beijing Academy of Quantum Information Sciences, Beijing 100193, China}
\affiliation{School of Physics and Key Laboratory of Quantum State Construction and Manipulation (Ministry of Education), Renmin University of China, Beijing 100872, China}
\author{Vasilii Smirnov}
\affiliation{P.N. Lebedev Physical Institute of the Russian Academy of Sciences, Moscow 119991, Russia}
\affiliation{Russian Quantum Center (RQC), Moscow 121205, Russia}
\author{Kaiye Shi}
\affiliation{Beijing Academy of Quantum Information Sciences, Beijing 100193, China}
\affiliation{School of Physics and Key Laboratory of Quantum State Construction and Manipulation (Ministry of Education), Renmin University of China, Beijing 100872, China}
\author{Wei Zhang}
\thanks{wzhangl@ruc.edu.cn}
\affiliation{School of Physics and Key Laboratory of Quantum State Construction and Manipulation (Ministry of Education), Renmin University of China, Beijing 100872, China}
\affiliation{Beijing Key Laboratory of Opto-electronic Functional Materials and Micro-nano Devices, Renmin University of China, Beijing 100872, China}
\affiliation{Beijing Academy of Quantum Information Sciences, Beijing 100193, China}

\begin{abstract}
Exceptional points featuring enhanced energy response to perturbation hold significant potential in detection and measurement of weak signals. Of particular interest is the existence and property of high-order exceptional points in quantum systems, owing to the capability to provide high-order response to perturbations. We investigate the exceptional points in a system of $n$ identical qubits possessing parity-time-reversal symmetry. We prove that owing to an incomplete coalescence of eigenstates, the highest possible order of exceptional point is $n+1$, which is also the upper bound of the order of energy response to perturbation. More interestingly, by considering an Ising-type interaction, we analytically prove that to achieve an $(m+1)$-th order response for any $m \le n$, the system must acquire a nontrivial $m$-body interaction. Finally, we propose a Floquet driving scheme to implement an effective multi-body Ising-type interaction, which can be realized in trapped ions or superconducting qubits.
\end{abstract}

\maketitle



\section{Introduction}
Exceptional points (EPs)~\cite{Kato-1966}, known as energetically degenerate points in company with simultaneous coalescence of eigenstates~\cite{Berry-2004,Heiss-2004,Rotter-2008,Rotter-2009,Heiss-2012,Ueda-2020}, have attracted a surge of fundamental and practical interests in various physical systems, including microcavities~\cite{Hodaei-2017,Chen-2017,zhong2020hierarchical}, coupled resonators~\cite{Ding-2016,Wang-2019,Peng-2014,xu2024robust}, electronic circuits~\cite{Schindler-2011,Sakhdari-2019,Jose-2023,zhao2024exceptional,chen2024ultra}, superconducting circuits~\cite{song2024experimental}, waveguides~\cite{Zhang-2019,Liu-2020}, optomechanical systems~\cite{Jing-2017,Djorwe-2019,Xiong-2021}, cold atoms~\cite{Li-2019,Ashida-2017}, thermal atomic ensembles~\cite{liang2023observation, zhang2024realizing}, single photons~\cite{Xue-2017}, nuclear spins~\cite{Long-2020}, nitrogen-vacancy centers~\cite{Du-2019,Du-2024}, quantum gases~\cite{pan2019high}, electronic fluids~\cite{aquino2024laser}, plasmons~\cite{kiselev2024inducing} and trapped ions~\cite{Ding-2021}. Potential applications have been widely explored in loss-induced transparency~\cite{Painter-2011}, energy transfer~\cite{Harris-2016}, unidirectional lasing~\cite{Feng-2017}, laser absorption~\cite{soleymani2022chiral,wong2016lasing} and enhanced sensing~\cite{Hodaei-2017,Chen-2017,wiersig2016sensors,de2022design,jiang2022enhanced,zhong2019sensing,farhat2020pt,park2020symmetry,jiang2022exceptional,li2021exceptional,zhang2022anti}.
Recent studies of this topic are closely related to parity-time-reversal ($\mathcal{PT}$) symmetric Hamiltonians~\cite{Ganainy-2018}, where EPs are intrinsically linked with the transition from $\mathcal{PT}$ symmetry preserved (PTS) regime with real spectra to broken (PTB) regime with complex conjugated energies~\cite{Bender-1998}.

One of the most prominent features of EPs is the fractional power dependence of eigenenergy upon a perturbation $\lambda\sim|\epsilon|^{1/n}$~\cite{Kato-1966,Berry-2004,Heiss-2004,Rotter-2008,Rotter-2009,Heiss-2012,Ueda-2020}, where $n$ denotes the response order. This is in stark contrast to a linear response in Hermitian systems around a degenerate point solely in energies, namely diabolic point (DP)~\cite{Berry-1984}. In principle, a high-order EP (HOEP) with $n$-fold degeneracy, denoted by EP$_n$, can boost nonlinear energy splitting up to $n$-th order by applying approximate external perturbation~\cite{Zou-2021}. This observation thus offers an appealing possibility of enhancing weak signals which are undetectable under linear response schemes, and has potential applications in gyroscope~\cite{Chow-1985,Sunada-2007}, magnetometer~\cite{Rondin-2014}, nanomechanical mass sensing~\cite{Santos-2010} and particle detection~\cite{Zhu-2010,He-2011,Vollmer-2012}. 
Up to now, HOEPs have been experimentally demonstrated in multiple platforms, including classical photonic and acoustic systems~\cite{Hodaei-2017,Ding-2016,Wang-2019}, thermal atoms~\cite{liang2023observation, zhang2024realizing}, ions~\cite{Ding-2021} and a quantum multi-qubit system of nitrogen-vacancy centers~\cite{Du-2024}.

Here, we consider a quantum many-body system composed by $n$ identical qubits with $\mathcal{PT}$ symmetry to clarify the possibility and condition to host HOEPs, and of particular interest, the most desired high-order response. In the absence of inter-qubit interaction, we discuss the order of EP and the coalescence of states and energies, and observe an incomplete coalescence of eigenstates by estimating trace distances between states. We then evaluate the order of response in the presence of Ising interaction of different types. The most prominent finding is that an $(m+1)$-th order response (with $m \le n$) can only be achieved when an interaction involving no less than $m$-body is present. That is, if one wish to realize the maximal $(n+1)$-th order enhancement of sensing in an $n$-qubit system, a highly nontrivial $n$-body interaction must be introduced. Finally, we propose to realize such multi-body interaction and observe high-order response of HOEP in periodically driven systems.

\section{HOEPs in a non-interacting $n$-qubit system} 
We start from a non-interacting model comprising $n$ identical qubits,
\begin{eqnarray}
\label{eq1}
\mathcal{H}_0 = \sum^n_{j=1} h_j,
\end{eqnarray}
where $h_j=J(\sigma_x^j + i\gamma \sigma_z^j)$ is the single-particle two-level Hamiltonian with $\sigma_{x,y,z}$ the Pauli operators, $J>0$ is the coupling between two levels and $\gamma\ge 0$ is the dimensionless dissipation rate. For brevity, we omit the two-dimensional identity operator $\sigma_0$. The single-qubit Hamiltonian $h_j$ possesses $\mathcal{PT}$ symmetry $[h_j, \mathcal{PT}]=0$, with parity operator $\mathcal{P} = \sigma_x$ and time-reversal operator $\mathcal{T}$ being complex conjugate. The two eigenvalues are $\lambda_{j,\pm}=\pm J\varepsilon_{0}$ with $\varepsilon_{0}=\sqrt{1-\gamma^2}$, and the corresponding eigenstates, denoted by $\ket{{\pm}}$, take the form $\ket{{\pm}} = \ket{x_{j,\pm}}=(i\gamma\pm\varepsilon_{0},1)^{\rm T}/\sqrt{2}$ in the eigenspace $\{\ket{\phi_{j,\uparrow}},\ket{\phi_{j,\downarrow}}\}$ of $\sigma_z^j$. When $0\le\gamma<1$, the eigenvalues are real and the qubits are in the PTS regime. In contrast, in the PTB regime of $\gamma>1$, the eigenvalues become purely imaginary and complex conjugated~\cite{Bender-1998}. The transition occurs at an EP$_2$ with $\gamma=1$, where both the two eigenstates and eigenvalues degenerate, that is $\lambda_{j,\pm}=0$ and $\ket{x_{j,\pm}}_{\gamma=1}=(i,1)^{\rm T}/\sqrt{2}\equiv\ket{x_0}$.
The single-qubit Hamiltonian can be mapped into a purely dissipative Hamiltonian $h_j^{\rm Diss}=h_j-i\gamma J\sigma_0$. It stands as a reasonable approximation to the Lindblad equation by ignoring the quantum jump, as if we investigate static behavior or dynamics when the dissipation is not strong and the evolution time is not long.

In the absence of inter-qubit interaction, the eigenvalues $\lambda$ of the non-interacting system can be expressed as the summation of individual single-particle energies, i.e., $\lambda=\sum_{j=1}^n\lambda_{j,\beta_j = \pm}$.  
The eigenstates, excluding those at the EP, can be explicitly expressed by the Kronecker product of single-particle eigenstates $\ket{\chi}=\otimes_{j=1}^n\ket{x_{j,\beta_j}}$, forming $2^n$ distinct eigenstates labeled by $\ket{\chi_k}$ with $k=1,2,\cdots,2^n$. These states form a complete basis denoted as $\ket{\vec{\chi}}=(\ket{\chi_1},\ket{\chi_2},\cdots,\ket{\chi_{2^n}})$. 
Therefore, the Hilbert space is well established by the subspaces spanned by the eigenstates $\ket{\chi_k}$ of a $C_n^m$-fold degeneracy, satisfying $\mathcal{H}_0\ket{\chi_k}=\lambda_m\ket{\chi_k}$ and $\lambda_m=(n-2m)J{\varepsilon}_0$. Among these subspaces, the one with energy $\lambda_m$ exhibits eigenvalue degeneracy but without the coalescence of eigenstates, akin to a DP in a Hermitian system~\cite{Berry-1984}.
Unfortunately, such a Kronecker product approach becomes invalid at the EP, where the Hilbert space collapses and the Hamiltonian matrix is non-diagonalizable. 
For instead, we can span the degenerate subspace with a given $\lambda=\lambda_m$ by using appropriate linear combinations of $\ket{\chi_k}$ that satisfy $\mathcal{H}_0\ket{\chi_k}=\lambda_m\ket{\chi_k}$ to prevent eigenstate coalescence. Details for constructing non-coalescing eigenstates are given in the Appendix~\ref{sec:MethodsI}. Although the eigenstates with different eigenvalues coalesce at the EP, the ones that are already degenerate away from EP remain linearly independent. Consequently, even when all $2^n$ eigenvalues coalesce at the EP, there persist $C_n^{\lfloor n/2\rfloor}$ linearly independent eigenstates, ensuring that the Hilbert space does not completely collapse. The resulting EP is a superposition of $[C_n^m-C_n^{m-1}\Theta(m-1/2)]$-fold EP$_{n-2m+1}$'s with $m=\{0,1,\cdots,\lfloor n/2\rfloor\}$, rather than a one-fold EP$_{2^n}$. Here, $\Theta(x)$ is the Heaviside step function.

\begin{figure}[t!]     
	\centering
	\includegraphics[width=0.9\linewidth]{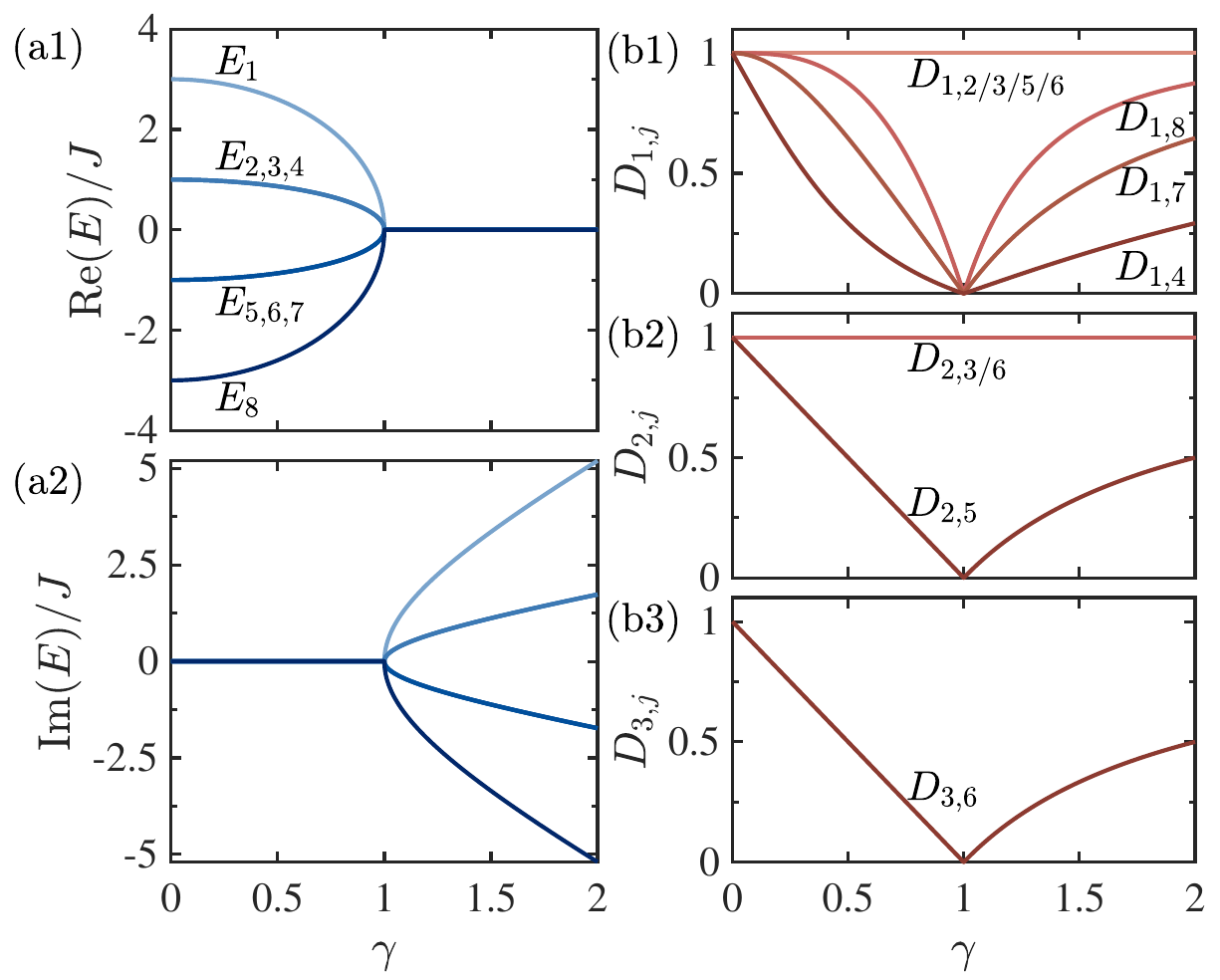}
	\caption{(a1) The real and (a2) imaginary parts of the energy spectrum for a non-interacting three-qubit system by varying the dissipation rate $\gamma$. Trace distance (b1) $D_{1,k}$ between $\ket{\psi_1}$ and $\ket{\psi_k}$ with $k\in\{2,3,\cdots,8\}$, (b2) $D_{2,k}$ between $\ket{\psi_2}$ and $\ket{\psi_k}$ with $k\in\{3,5,6\}$, (b3) $D_{3,6}$ between $\ket{\psi_3}$ and $\ket{\psi_6}$ are shown as functions of $\gamma$.
	}
	\label{fig1} 
\end{figure}

To illustrate the incomplete coalescence of eigenstates, we take a system of three qubits as an example. We sort the eight eigenvalues in descending order and label them by $E_{k=1,\dots,8}$ with corresponding eigenstates $\ket{\psi_k}$. The highest and lowest eigenvalues $E_{1,8}=\pm 3J{\varepsilon}_0$ have 1-fold degeneracy, while $E_{2,3,4}=J{\varepsilon}_0$ and $E_{5,6,7}=-J{\varepsilon}_0$ are 3-fold degenerate. All eigenvalues become degenerate to zero at $\gamma=1$, as shown in Fig.~\ref{fig1}(a). The coalescence of states can be captured by the trace distance between two states, defined as~\cite{Nielsen-2000} 
\begin{eqnarray}
D_{k,l}=\frac{1}{2}{\rm tr}|{\pmb \rho}_k-{\pmb \rho}_l|
\end{eqnarray}
with $|{\pmb \rho}|=\sqrt{{\pmb \rho}^{\dag}{\pmb \rho}}$ and ${\pmb \rho}_k=\ket{\bar{\psi}_k}\bra{\bar{\psi}_k}$ the density matrix of the normalized state $\ket{\bar{\psi}_k}=\ket{\psi_k}/\sqrt{\bra{\psi_k}\psi_k}\rangle$. Two states are maximally distanced (orthogonal) when $D_{k,l}=1$, while they are identical except for a phase difference when $D_{k,l}=0$~\cite{Ueda-2017,Guo-2020}. 
In Fig.~\ref{fig1}(b1), we present the trace distance $D_{1,k}$ between $\ket{\psi_1}$ with the highest energy $3J\varepsilon_0$ and the others by varying the dissipation $\gamma$. Four of them, $D_{1,k\in\{2,3,5,6\}}$, become equal to one at $\gamma=1$, ensuring that $\ket{\psi_1}$ is orthogonal to $\ket{\psi_{k\in\{2,3,5,6\}}}$ at the EP. In contrast, the other three trace distances, $D_{1,k\in\{4,7,8\}}$, fall into zero and signify the coalescence of $\ket{\psi_1}$ and $\ket{\psi_{k\in\{4,7,8\}}}$.

Then we evaluate the trace distance between $\ket{\psi_2}$ and $\ket{\psi_{k\in\{3,5,6\}}}$ in Fig.~\ref{fig1}(b2). While $D_{2,k\in\{3,6\}}$ remain unity, $D_{2,5}$ depends on dissipation and drops to zero at the EP, indicating the coalescence of $\ket{\psi_2}$ and $\ket{\psi_{5}}$. Lastly, we calculate $D_{3,6}$ in Fig.~\ref{fig1}(b3) and find that $\ket{\psi_3}$ and $\ket{\psi_6}$ coalesce at the EP. 
These results demonstrate that the EP of a three-qubit system is a combination of 1-fold EP$_4$ and a 2-fold EP$_2$, acquiring three linearly independent states at this critical point. Such three states cannot be constructed by the Kronecker product of single-particle states. For instead, as shown in Appendix~\ref{sec:MethodsI}, we need to work in the eigenspace of total spin operator to obtain 
\begin{eqnarray}
\ket{\eta_1} &=& \ket{+++}_{\gamma=1}, 
\nonumber \\
\ket{\eta_2} &=& \frac{1}{\sqrt{2}{\varepsilon}_0}\big(\ket{++-}-\ket{-++}\big)_{\gamma=1},
\nonumber \\
\ket{\eta_3} &=& \frac{1}{\sqrt{6}{\varepsilon}_0}\big(\ket{++-}+\ket{-++}-2\ket{+-+}\big)_{\gamma=1}.
\end{eqnarray}
%


\section{High-order response induced by interaction.} 
One of the most intriguing properties of HOEP is that the enhanced response to external perturbation $\epsilon$, scaling as $\epsilon^{1/p}$ with the response order $p$, can attain the order of EP~\cite{Zou-2021}. This implies that an $n$-qubit system can achieve a response up to $(n+1)$-th order upon perturbation. Next, we consider a weak Ising-type interaction $\mathcal{H}_{\rm int}=\epsilon\otimes_{j=1}^n\sigma_{k_j}^j$ with ${k_j\in\{0,x,y,z\}}$, and demonstrate that an $(n+1)$-th order response can only be witnessed by a nontrivial $n$-body interaction. More generally, we will prove that the maximal response order one can get in a system with interaction involving $m$ ($m<n$) bodies is $m+1$. These findings suggest that HOEP is an anisotropic EP~\cite{Chan-2018,Chan-2019}, showing different response behavior when it is approached along distinct directions in the parameter space. In the following discussion, we adopt the notation $\mathcal{H}_0\equiv\mathcal{H}_0(\gamma=1)$ for simplification, and take $J=1$ as the energy unit unless otherwise specified.

Before proceeding, we notice that the eigenequation for $\mathcal{H}_0$ at EP simplifies to
\begin{equation}
	\lambda^{2^n}=0.
	\label{h0}
\end{equation}
In the presence of a perturbation $\epsilon$, the eigenequation for the total Hamiltonian $\mathcal{H}=\mathcal{H}_0+\mathcal{H}_{\rm int}$ can be written as polynomial of order $2^n$, with the coefficient of the leading term $(\lambda^{2^n})$ being free of $\epsilon$:
 \begin{equation}
 	\lambda^{2^n}+\sum_{i=0}^{2^n-1}\alpha_{i}(\epsilon)\lambda^i=0,
 \end{equation}
 %
with $\alpha_i(\epsilon)$ - polynomial of $\epsilon$. We neglect all terms containing $e^{x}$ ($x \geq 2$) given that $\epsilon$ is small. Then the eigenequation takes the form
 \begin{equation}
 	\lambda^{2^n}+\sum_{i=k}^{2^n-1}\,\beta_{i}\,\epsilon\lambda^{i}
	 = \lambda^k(\lambda^{2^n-k}+...+\beta_{k}\,\epsilon) = 0.
 	\label{eigen}
 \end{equation}  
 %
 Here $k$ is the lowest power of $\lambda$ among terms linear by $\epsilon$. It can be seen from the bracket in (\ref{eigen}) that the response order will be $2^n-k$. Thus, we assert that to induce the maximal $(n+1)$-th order response in an $n$-qubit system, the eigenequation of the perturbed Hamiltonian $\mathcal{H}$
 \begin{equation}
 	\lambda^{2^n-n-1}(\lambda^{n+1}+...+\beta_{2^n-n-1}\epsilon) = 0
 \end{equation}
must encompass a special term in the form of $\epsilon\lambda^{2^n-n-1}$.

Then, we prove that an Ising-type interaction involving only $m$ bodies with $m<n$ fails to induce an $(n+1)$-th order response. To begin with, we consider an $m$-body Ising-type interaction 
\begin{equation}
	\mathcal{H}_{\rm int}^{m}=\epsilon\otimes_{j=1}^m\sigma_{k_j}^j
\end{equation}
with intensity $\epsilon$ applied to a non-interacting $n$-qubit system described by 
\begin{equation}
	\mathcal{H}_0=\mathcal{H}_0^{m}+\mathcal{H}_0^{n-m},
\end{equation} 
where $\mathcal{H}_0^{m}=\sum_{j=1}^mh_j$ and $\mathcal{H}_0^{n-m}=\sum_{j=m+1}^nh_j$. Notice that by far, the interaction is only introduced among the first $m$ qubits, while the rest $(n-m)$ ones remain uncoupled. The total Hamiltonian can be expressed as 
\begin{equation}
	\mathcal{H}=(\mathcal{H}_0^m+\mathcal{H}_{\rm int}^m)\otimes\mathbb{I}_{n-m}+\mathbb{I}_m\otimes\mathcal{H}_0^{n-m}
\end{equation}
with $\mathbb{I}_{m}=\otimes_{j=1}^m\sigma_0^j$ and $\mathbb{I}_{n-m}=\otimes_{j=m+1}^n\sigma_0^j$, which can be diagonalized by the unitary operator $\mathcal{U}=\mathcal{U}_m\otimes\mathcal{U}_{n-m}$, such that 
 \begin{equation}
 	\mathcal{U}^{-1}\mathcal{H}\,\mathcal{U}=\mathcal{U}^{-1}[(\mathcal{H}_0^m+\mathcal{H}_{\rm int}^m)\otimes\mathbb{I}_{n-m}]\,\mathcal{U}.
\end{equation}
 Here, we have used the fact that 
 \begin{equation}
 	[(\mathcal{U}_{n-m})^{-1}\mathcal{H}_0^{n-m}\mathcal{U}_{n-m}]_{\gamma\to 1}=0.
\end{equation} 
 The diagonalization can be converted into finding the solution of the eigenequation 
 \begin{equation}
 	|\mathcal{H}-\lambda\mathbb{I}_n|=|(\mathcal{H}_0^m+\mathcal{H}_{\rm int}^m-\lambda\mathbb{I}_{m})\otimes\mathbb{I}_{n-m}|=0.
\end{equation} 
By resorting to the formula $|A\otimes B| = |A|^m \cdot |B|^n$~\cite{Horn-1991} with an $n\times n$ matrix $A$ and an $m\times m$ matrix $B$, one obtains 
 \begin{equation}
 	|(\mathcal{H}_0^m+\mathcal{H}_{\rm int}^m-\lambda\mathbb{I}_{m})\otimes\mathbb{I}_{n-m}|=|\mathcal{H}_0^m+\mathcal{H}_{\rm int}^m-\lambda\mathbb{I}_{m}|^{2^{n-m}},
\end{equation}
which effectively reduces the eigenequation of an $n$-qubit system to that of an $m$-qubit system.
In the eigenequation of the $m$-qubit system, the lowest possible power of $\lambda$ among the terms proportional to $\epsilon$ is $2^m-m-1$, since the highest possible response order is EP order $m+1$. Thus, we can immediately conclude that the lowest power of $\lambda$ in all terms proportional to $\epsilon$ in $|\mathcal{H}-\lambda\mathbb{I}_n|$ is not less than $2^n-m-1$, such that an $m$-body Ising-type interaction can induce an $(m+1)$-th order response at most.

For a general, and from an experimental perspective, more realistic case where more than one interaction term is present, i.e., $\mathcal{H}_{\rm int}=\sum_i\mathcal{H}_{\rm int}^{m_i}$ with $m_i<n$ for all $i$, the term in linear dependence on $\epsilon$ in the eigenequation can be represented by a linear combination of those when considering these interactions individually, as one would naturally expect for a first-order perturbation. 
A precise proof of the summation rule for terms proportional to $\epsilon$ is demonstrated in Appendix~\ref{sec:MethodsII}. As a result, the power of $\lambda$ in the special term governed by $\mathcal{H}_{\rm int}$ is no less than the lowest power of $\lambda$ in the corresponding terms when $\mathcal{H}_{\rm int}^{m_i}$ is present individually. Therefore, we conclude that an $m$-body Ising-type interaction with $m<n$ cannot induce an $(n+1)$-th order response.

\begin{figure}[!t]
	\centering
	\includegraphics[width= 0.95\linewidth]{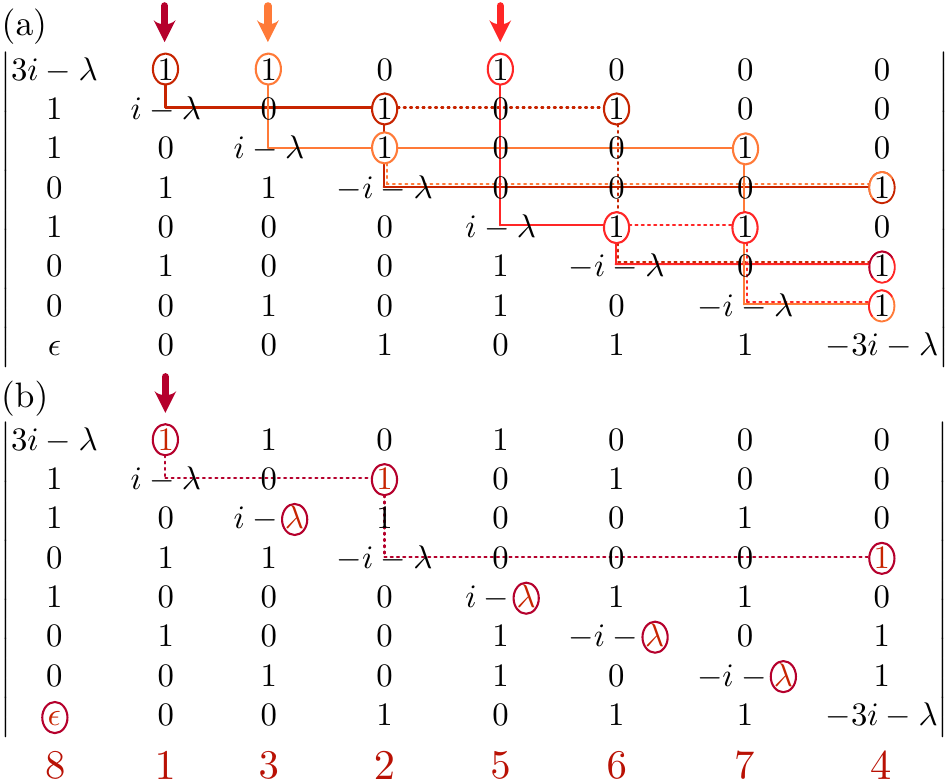}
\caption{(a) Permutation rule for the special term in eigenequation. A three-qubit system with an interaction $\mathcal{H}_0^3+\mathcal{H}_s^{3,1}$ is considered. Each permutation is denoted by a line starting from a nonzero off-diagonal element in the first row (denoted by colored arrows) and ending at a nonzero element in the last column. (b) For one example permutation, the term is composed by all circled elements. The sign of the permutation (e.g., 81325674) can be obtained from off-diagonal elements only, leading to $(-1)^n$.}
\label{fig2}
\end{figure}

Then, we discuss the case of $n$-body Ising-type interaction by considering an example 
\begin{equation}
	\mathcal{H}_{\rm int}^n=\epsilon\otimes_{j=1}^{2^n}\sigma_x^j,
\end{equation} 
which can be implemented in trapped ions~\cite{Katz-2023,Katz-2022} or superconducting qubits~\cite{Scully-2022}.
This example serves as a handy candidate to demonstrate the capability for generating the $(n+1)$-th order response around EP. In the basis of $\{ \otimes_{j=1}^n \vert \phi_{j,\uparrow/\downarrow} \rangle \}$, the interaction can be represented in a matrix form with off-diagonal elements only, which are proportional to $\epsilon$. When analyzing the leading order of response induced by the interaction, we can split the interaction part into $2^n$ matrices holding a single non-zero element $\epsilon$, consider the arising special terms one by one and then find the maximal response order according to summation rule in Appendix~\ref{sec:MethodsII}. For example, if the interaction presents an element in the left-bottom corner (as demonstrated in Fig.~\ref{fig2} for a three-qubit system), which can be written as $\mathcal{H}_s^{n,1}$ with the elements $(\mathcal{H}_s^{n,1})_{k,l}=\epsilon\,\delta_{k,2^n}\delta_{l,1}$, the eigenequation for the total Hamiltonian $\mathcal{H}^n_0+\mathcal{H}_s^{n,1}$ reads
\begin{equation}
|\mathcal{H}_0^n+\mathcal{H}_{s}^{n,1}-\lambda \mathbb{I}| = \lambda^{2^n}-n!\,\epsilon\,\lambda^{2^n-n-1}.
\label{eq:Hs}
\end{equation}
%

To see this, we first notice that since the maximal EP order is $n+1$, all terms with the power of $\lambda$ lower than $2^n-n-1$ are prohibited. 
Now we show that the special term proportional to $\epsilon$ is contributed by permutations involving $n+1$ off-diagonal matrix elements, including the one of $\epsilon$ at the left-bottom corner and others of value unity. 
Such permutations can be obtained by following a specific pattern as depicted in Fig.~\ref{fig2}(a): start from a non-zero off-diagonal element in the first row, go downwards and make a left turn upon encountering a diagonal element, and make a right turn upon encountering another non-zero off-diagonal element, and so on. The resultant permutations (example in Fig.~\ref{fig2}(b)) consist of $n$ off-diagonal elements (circled in red) of the path (dashed), together with the $\epsilon$ element sitting at the left-bottom corner and the $2^n-n-1$ diagonal elements in unchosen rows or columns (also circled in red). Here we remind that all terms with the power of $\lambda$ lower than $2^n-n-1$ are prohibited and therefore will automatically cancel each other in the resulting eigenequation, therefore we only consider picking $\lambda$ from the diagonal matrix elements. Each permutation path contributes a term with one power of $-\lambda$ from each diagonal matrix element, and a constant of unity or $\epsilon$ from each non-diagonal element, leading to $\epsilon\,(-\lambda)^{2^n-n-1}$. 
In addition, the sign contributed by the inversion number of every permutation [e.g., (81325674) in Fig~\ref{fig2}(b)] is $(-1)^{n}$, such that the total sign before $\epsilon\lambda^{2^n-n-1}$ is always negative. Taking into account Eq.~(\ref{h0}), this procedure is essentially the decomposition of determinant along rows, therefore no other terms will be present.
Lastly, by adding an extra qubit to an $(n-1)$-qubit system, the matrix Hamiltonian will be doubled in dimension, with a duplicated diagonal block and two off-diagonal blocks. Since the appended qubit has interaction with all $(n-1)$-qubits, the off-diagonal blocks have one nonzero element in each row. Then the total number of permutations is increased by $n$ times, leading to the prefactor of $n!$ for the special term in Eq.~(\ref{eq:Hs}).

In similar manners, Eq.~(\ref{eq:Hs}) applies to $\mathcal{H}_{s}^{n,j}$ with a single $\epsilon$ placed at an arbitrary position along the side diagonal, formally expressed as
\begin{equation}
	(\mathcal{H}_s^{n,j})_{k,l}=\epsilon\,\delta_{k,2^n+1-j}\delta_{l,j},
\end{equation}
where $j\in\{1,2,\cdots,2^n\}$. 
By employing the summation rule, the special term for an $n$-body Ising-type interaction
\begin{equation}
	\mathcal{H}_{\rm int}^n=\epsilon\otimes_{j=1}^{2^n}\sigma_x^j,
\end{equation}
further written as
\begin{equation}
	\mathcal{H}_{\rm int}^n=\sum_{j=1}^{2^n}\mathcal{H}_s^{n,j},
\end{equation}
is the direct summation of the special terms for each individual $\mathcal{H}_s^{n,j}$. Thus, the term in the eigenequation being proportional to $\epsilon$ and having the lowest order of $\lambda$ is $-2^n n!\,\epsilon\lambda^{2^n-n-1}$. Since the maximum EP order of an $n$-qubit system is $n+1$, the terms lower than $\lambda^{2^n-n-1}$ in the total eigenequation is proportional to at least $\epsilon^2$, which can be neglected.
The eigenequation is thus approximated as 
\begin{equation}
	|\mathcal{H}_0^n+\mathcal{H}_{\rm int}^n-\lambda\mathbb{I}_n|\approx\lambda^{2^n}+\cdots-2^n n!\,\epsilon\lambda^{2^n-n-1},
\end{equation} 
with solutions acquired $(n+1)$-th order enhanced response
\begin{equation}
	\lambda \propto \epsilon^{1/(n+1)}.
\end{equation}
%

\begin{figure}[!t]
	\centering
	\includegraphics[width=0.98\linewidth]{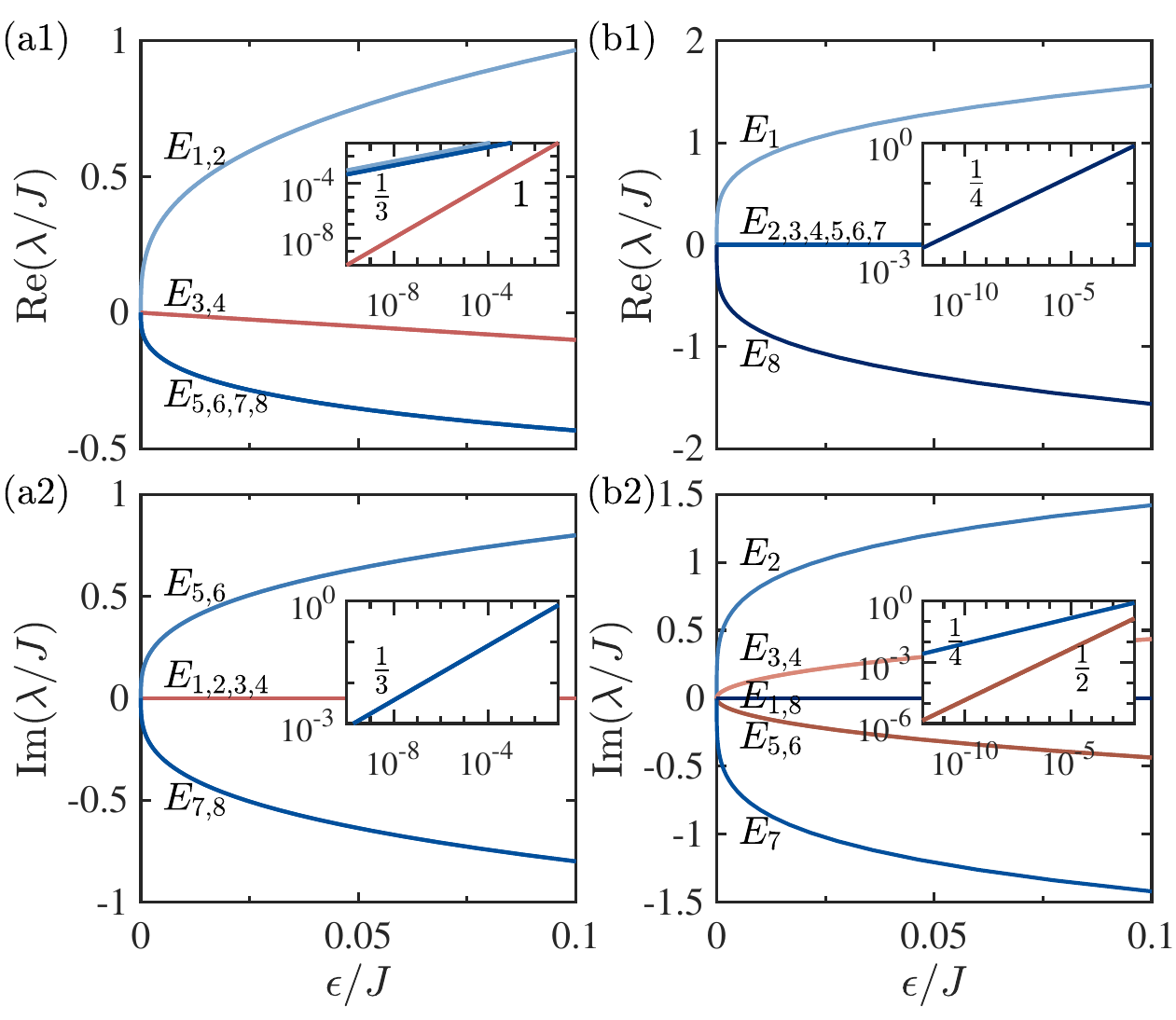}
	\caption{(a1) The real and (a2) imaginary parts of energy response to a two-body Ising-type interaction $H_{\rm int}=\epsilon\,\sigma_{x}^1\sigma_{x}^2\sigma_0^3$ in a three-qubit system. (b1) and (b2) depict the corresponding results for a three-body Ising-type interaction $H_{\rm int}=\epsilon\,\sigma_{x}^1\sigma_{x}^2\sigma_x^3$. The insets display the modulus of energy in a logarithmic scale.
	}
	\label{fig3} 
\end{figure}

To illustrate the effect of interaction, we consider a three-qubit system and plot the eigenenergies in the presence of a two-body interaction [Fig.~\ref{fig3}(a)]
\begin{equation}
	\mathcal{H}_{\rm int}=\epsilon\,\sigma_{x}^1\sigma_{x}^2\sigma_0^3
\end{equation} 
or a three-body interaction [Fig.~\ref{fig3}(b)]
\begin{equation}
	\mathcal{H}_{\rm int}=\epsilon\,\sigma_{x}^1\sigma_{x}^2\sigma_x^3.
\end{equation}
The insets in all panels display the energies in a logarithm scale. One can clearly see that while a third-order response can be induced by a two-body interaction, as evidenced by the $1/3$ slope in the insets of Figs.~\ref{fig3}(a), an enhanced response to the fourth order is achieved with a $1/4$ slope in the inset of Fig.~\ref{fig3}(b) for the case of a three-body interaction.

\section{Experimental realization in Floquet systems} 
\begin{figure}[t!]     
	\centering
	\includegraphics[width=0.98\linewidth]{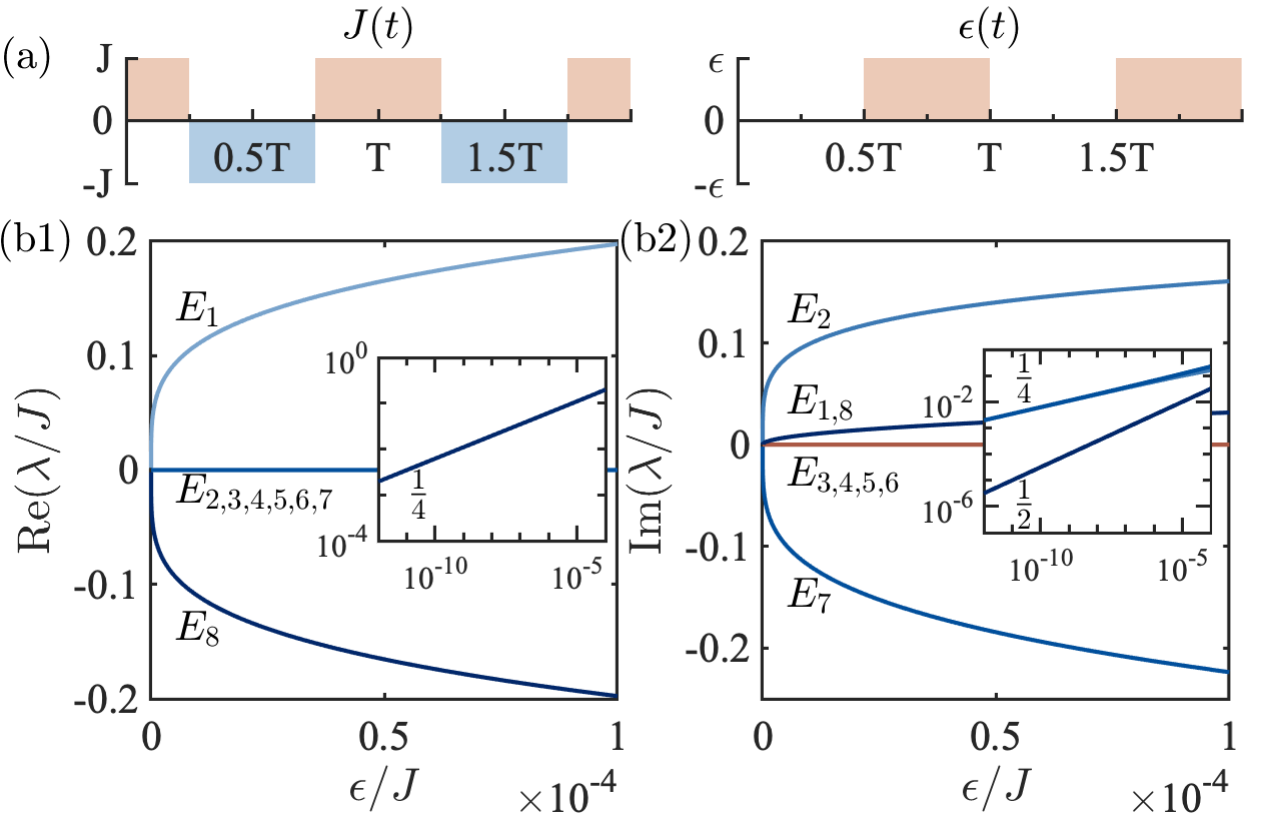}
	\caption{(a) The periodic driving protocol to realize an effective three-body interaction. (b1) and (b2) show the real and imaginary components of the energy spectrum corresponding to the effective Hamiltonian  Eq.~(\ref{Heff}) around an EP$_4$ as functions of the perturbation strength $\epsilon$. In these plots, we set $TJ=0.2$. 
	}
	\label{fig4} 
\end{figure}
Finally, we propose to realize a multi-body interaction by periodically driving a Hamiltonian with two-body interaction 
\begin{eqnarray}
\mathcal{H}(t) = J(t)\bar{\mathcal{H}}_1 + \epsilon(t)\bar{\mathcal{H}}_2 + \mathcal{H}_0,
\end{eqnarray}
where the two interaction terms are
\begin{eqnarray}
\bar{\mathcal{H}}_1 = \sum_{k\neq l}\sigma_{x}^k\sigma_{y}^l,\quad\bar{\mathcal{H}}_2 = \sum_{k\neq l}\sigma_{z}^k\sigma_{x}^l.
\end{eqnarray}
The parameters are assumed to vary in the form of step functions within a driving period $T$, as illustrated in Fig.~\ref{fig4}(a),
\begin{eqnarray}
J(t) &=& J\Big[\Theta(\frac{T}{4}-t) + \Theta(T-t)\Theta(t-\frac{3}{4}T) \notag\\
&&\quad- \Theta(t-\frac{T}{4})\Theta(\frac{3}{4}T-t)\Big],
\\
\epsilon(t) &=&\epsilon\,\Theta(t-\frac{T}{2}),
\end{eqnarray}
where $\Theta(x)$ is the Heaviside step function.
The corresponding time-evolution operator with the aid of Magnus expansion~\cite{Kennes-2020} at high frequency ($1/T\gg \epsilon, J, \gamma$) can be written as $\mathcal{U}(t) = {\rm exp}[{\Omega(t)}] = {\rm exp}[{\sum_{k=0}^{+\infty}\Omega_k(t)}]$, where 
\begin{eqnarray}
\Omega_1(t) &=& \int_0^tdt_1 A(t_1),
\nonumber \\
\Omega_2(t) &=& \frac{1}{2}\int_0^tdt_1\int_0^{t_1}dt_2[A(t_1),A(t_2)]
\end{eqnarray}
with $A(t)=-i\mathcal{H}(t)$. At the time of complete periods, the evolution can be described by an effective Floquet Hamiltonian $\mathcal{H}_{\rm eff} = i(\Omega_1+\Omega_2+\cdots)/T$.
For a three-qubit system, the effective Hamiltonian up to the second-order expansion is given by 
\begin{eqnarray}
\mathcal{H}_{\rm eff}
= \Big(\mathcal{H}_0 + \frac{1}{2}\epsilon \bar{\mathcal{H}}_2\Big) +i\epsilon\frac{T}{8}\Big([\mathcal{H}_0,\bar{\mathcal{H}}_2]+\frac{1}{2}J[\bar{\mathcal{H}}_2,\bar{\mathcal{H}}_1]\Big).\notag\\
\label{Heff}
\end{eqnarray}
An effective three-body interaction thus emerges in the commutation operator $[\bar{\mathcal{H}}_2,\bar{\mathcal{H}}_1]$.
The effective Hamiltonian Eq.~(\ref{Heff}) contains both two- and three-body interactions, and presents fourth-order response to the interaction strength $\epsilon$ around the EP$_4$, as witnessed by the 1/4 slope in the inset of Figs.~\ref{fig4}(b). An effective interaction involving more particles can in principle be realized by going to higher orders of Magnus expansion.

\section{Summary} 
We study high-order exceptional points (EPs) emerged in systems of $n$ identical qubits and the enhanced energy response to perturbations around the EP. In the absence of inter-qubit interaction, we find that the EP is superposed by a set of EPs with the highest order of $n+1$, attributed to the incomplete coalescence of eigenstates evidenced by the trace distances. More importantly, we conclude that even at a high-order EP, the energetic response to perturbation is not necessarily of the same order, but relies on the presence and specific form of inter-qubit interaction. For Ising-type interactions, we prove that one must impose an $m$-body interaction to induce an $(m+1)$-th order response. That is, to achieve the maximal $(n+1)$-th order enhanced response in an $n$-qubit system, a nontrivial $n$-body interaction must be invoked. This result shows the dependence of the sensitivity at exceptional points on the applied Ising perturbation, and can be extended to other types of interaction by decomposing it to Ising-type interactions and utilizing the summation rule (see Appendix~\ref{sec:MethodsII}). Finally, we propose to simulate an effective multi-body Ising-type interaction in periodically driven systems, which can be implemented with trapped ions. This allows to further investigate the behavior of the system under manually controlled perturbation and estimate the effects of external noise or decoherence for future applications. Our results set strict condition for the application of high-order EPs in quantum sensing and metrology.

\acknowledgments
This work is supported by the National Key R\&D Program of China (Grants No.~2022YFA1405301), and the National Natural Science Foundation of China (Grants No.~92265208 and No.~12074428). T.S. thanks support from the Postdoctoral Fellowship Program of CPSF (Grants No. GZC20232945).

\appendix

\section{Construction of non-coalescing eigenstates at exceptional points}
\label{sec:MethodsI}

A system of $n$ identical non-interacting qubits described by Eq.~(1) of the main text supports $n+1$ subspaces. Each of the subspaces is characterized by a $C_n^m$-fold degenerate eigenvalue $\lambda=(n-2m)J{\varepsilon}_0$, with $m\in\{0,1,\cdots,n\}$ and ${\varepsilon}_0=\sqrt{1-\gamma^2}$. The system has an exceptional point (EP) located at $\gamma=1$, where all eigenvalues become degenerate. At positions other than the EP, the eigenstates can be expressed in terms of the Kronecker products of single-particle eigenstates $\ket{\vec{\chi}}=(\ket{\chi_1},\ket{\chi_2},\cdots,\ket{\chi_{2^n}})$, where $\ket{\chi_{k}}=\otimes_{j=1}^n\ket{x_{j,\pm}}$ and the single-particle eigenstates $\ket{x_{j,\pm}} = (i\gamma\pm\varepsilon_0,1)^{\rm T}/\sqrt{2}$ represented in the basis $\ket{\phi_{j,\{\uparrow, \downarrow\}}}$ of the Pauli operator $\sigma_{z}$ acting on the $j$-th qubit. Thus, we can write the eigenstates of the system directly in terms of $\ket{\phi_{j,\{\uparrow, \downarrow\}}}$,
\begin{eqnarray}
\ket{\vec{\chi}}= U_{\chi} \ket{\vec{\Phi}},
\label{PhiToChi}
\end{eqnarray}
where $U_{\chi}$ is an invertible matrix, and $\ket{\vec{\Phi}}=(\ket{\Phi_1},\ket{\Phi_2},\cdots\ket{\Phi_{2^n}})$ with $\ket{\Phi_{k}}=\otimes_{j=1}^n\ket{\phi_{j,\alpha_j = \uparrow, \downarrow}}$. Notice that $\ket{\Phi_{k}}$'s form a basis for the total spin projection operator $\mathcal{S}_z\equiv\sum_{j=1}^n\sigma_z^j/2$ with eigenvalue $S_z=\sum_{j=1}^n\bar{\alpha}_j/2$, where $\bar{\alpha}_j=1$ for $\alpha_j=\uparrow$ and $\bar{\alpha}_j=-1$ for $\alpha_j=\downarrow$.
However, at the EP the system is non-diagonalizable and the eigenstates $\ket{\vec{\chi}}$ obtained from the Kronecker product of the single-particle eigenstates $\ket{x_{j,\pm}}$ will simultaneously coalesce, i.e. $\ket{\vec{\chi}}_{\gamma=1}=(\ket{\chi_0},\cdots,\ket{\chi_0})$ with $\ket{\chi_0}=\otimes_{j=1}^n\ket{x_0}$ and $\ket{x_0}=(i,1)^{\rm T}/\sqrt{2}$. On the other hand, by evaluating the trace distances between eigenstates, we find that they do not always evolve to zero as approaching EP. This observation suggests that the Kronecker product states can not be directly used to construct the non-coalescing eigenstates at the EP. For instead, next we show that one must span the $C_n^m$-dimensional subspaces by the eigenstates of collective spin operators, and use them to reach all non-coalescing eigenstates at the EP. In the following, we focus on the subspaces with non-negative energies, i.e. $m\le\lfloor n/2\rfloor$, and note that the negative counterpart can be obtained trivially by symmetry argument.

For a system of non-interacting qubits, the Hamiltonian $\mathcal{H}_0$ can be spanned by the common eigenstates of the total spin operator $\mathcal{S}^2\equiv(\sum_{j=1}^n\vec{s}_j)^2$ with $\vec{s}_j=(\sigma_x^j,\sigma_y^j,\sigma_z^j)/2$, and its projection operator $\mathcal{S}_z$. Since $\mathcal{H}_0$ commutes with $\mathcal{S}^2$, its matrix representation is block diagonal as 
\begin{eqnarray}
\bar{H}_0 =
\begin{pmatrix}
	(\bar{H}_0)_1& 0 &\cdots & 0\\
	0&(\bar{H}_0)_2&\cdots&0\\
	\vdots&\vdots&\ddots&\vdots\\
	0&0&\cdots&(\bar{H}_0)_{\lfloor\frac{N}{2}\rfloor+1}
\end{pmatrix}.
\label{barH0}
\end{eqnarray}
Here, the $k$-th diagonal block $(\bar{H}_0)_k$ corresponds to the subspace with a total spin of $S=n/2-k+1$, and a dimension of ${D}_k =  (2S+1)(C_n^{k-1}-C_n^{k-2}\Theta(k-3/2))$ with $\Theta(x)$ the Heaviside step function. Then, we can diagonalize the matrix $\bar{H}_0$ via block diagonalization. For the $k$-th block $(\bar{H}_0)_k$, there are $2S+1=n-2k+3$ different eigenenergies $\lambda_{j\in\{0,1,\cdots,2S\}}=2(S-j)J{\varepsilon}_0$, which are all degenerate of $d_k=(C_n^{k-1}-C_n^{k-2}\Theta(k-3/2))$ folds. Thus, the total dimension of $(\bar{H}_0)_k$ is $(n-2k+3)d_k$. When approaching the EP, the eigenstates corresponding to different eigenenergies will coalesce. However, since the spin symmetry presented in the degenerate subspace with the same eigenenergy $\lambda_j$ remains untouched, the dimension of such subspaces are kept at  $d_k$. As a consequence, the block $(\bar{H}_0)_k$ contributes a $d_k$-fold EP with EP order $2S+1=n-2k+3$. The $d_k$ non-coalescing eigenstates can be obtained by diagonalizing the degenerate subspace with the same eigenenergy $\lambda_j$ away from the EP, and setting $\gamma=1$ afterwards.

To gain a clear understanding of this procedure, we take an $n=3$-qubit system as an example, with Hamiltonian
\begin{eqnarray}
\mathcal{H}_0=h_1+h_2+h_3.
\end{eqnarray}
Away from the EP, the 8-dimensional matrix Hamiltonian can be written in the block form 
\begin{eqnarray}
\bar{H_0}=\begin{pmatrix}
	(\bar{H}_0)_1&0\\
	0&(\bar{H}_0)_2
	\end{pmatrix},
\end{eqnarray}
where
\begin{eqnarray}
	(\bar{H_0})_1=\begin{pmatrix}
		3iJ\gamma&\sqrt{3}J&0&0\\
		\sqrt{3}J&iJ\gamma&2J&0\\
		0&2J&-iJ\gamma&\sqrt{3}J\\
		0&0&\sqrt{3}J&-3iJ\gamma
	\end{pmatrix},
\end{eqnarray}
and 
\begin{eqnarray}
	(\bar{H_0})_2=\begin{pmatrix}
		iJ\gamma&0&-J&0\\
		0&iJ\gamma&0&-J\\
		-J&0&-iJ\gamma&0\\
		0&-J&0&-iJ\gamma
	\end{pmatrix}.
\end{eqnarray}
The first block $(\bar{H}_0)_1$ corresponds to the choice of $S=3/2$, and has four non-degenerate eigenenergies $\lambda_{j\in\{0,1,2,3\}}=2(S-j)J{\varepsilon}_0$ corresponding to four linearly independent eigenstates. When $\gamma=1$, all the four eigenstates coalesce to the same state, indicating a fourth-order EP. The second block $(\bar{H}_0)_2$ is spanned by states with $S=1/2$, and has two different eigenvalues $\lambda_{4,5} = J \varepsilon_0$ and $\lambda_{6,7} = -J \varepsilon_0$, both have a double-fold degeneracy. When approaching the EP, the eigenenergies $\lambda_{4,5,6,7}$ are degenerate, and the eigenstates associated with different $\lambda$ coalesce. However, within a same degenerate subspace, the two eigenstates $\ket{\psi_4}=\ket{\psi_6}$ and $\ket{\psi_5}=\ket{\psi_7}$ are still linearly independent, leading to a two-fold EP$_2$.

\section{Summation rule of terms proportional to $\epsilon$ in an eigenequation}
\label{sec:MethodsII}

Without loss of generality, we consider two Ising-type interactions, denoted by $\mathcal{H}_{\rm int,1}$ and $\mathcal{H}_{\rm int,2}$ in an $n$-qubit system, which are both in the form of $\epsilon\otimes_{j=1}^n\sigma_{k_j}^j$ where $k_j\in\{0,x,y,z\}$. By individually applying these interactions and their combination to the non-interacting Hamiltonian $\mathcal{H}_0$, the resulting total Hamiltonians are $\mathcal{H}_l=\mathcal{H}_0+\mathcal{H}_{{\rm int},l}$, where $l\in\{1,2,3\}$ and $\mathcal{H}_{\rm int,3} = \mathcal{H}_{\rm int,1}+\mathcal{H}_{\rm int,2}$. The terms proportional to $\epsilon$ in their corresponding eigenequations are denoted by $C_l\equiv\frac{\partial}{\partial\epsilon}{\rm det}(\mathcal{H}_l-\lambda\mathbb{I}_{n})|_{\epsilon=0}\cdot\epsilon$. Using the formula ${\rm det}A={\rm exp}[{\rm Tr}({\rm ln}A)]$, we obtain
\begin{eqnarray}
C_l &=& \frac{\partial}{\partial\epsilon}{\rm det}(\mathcal{H}_l-\lambda\mathbb{I}_{n})\Big|_{\epsilon=0}\cdot\epsilon\notag\\
&=& \frac{\partial}{\partial\epsilon}e^{{\rm Tr}\big[ {\rm ln}(\mathcal{H}_l-\lambda\mathbb{I}_n)\big]}\Big|_{\epsilon=0}\cdot\epsilon\notag\\
&=& {\rm det}(\mathcal{H}_l-\lambda\mathbb{I}_n)\, {\rm Tr}\Big[(\mathcal{H}_l-\lambda\mathbb{I}_n)^{-1}\frac{\partial\mathcal{H}_l}{\partial\epsilon}\Big]\Big|_{\epsilon=0}\cdot\epsilon\notag\\
&=& {\rm det}(\mathcal{H}_0-\lambda\mathbb{I}_n)\, {\rm Tr}\Big[(\mathcal{H}_0-\lambda\mathbb{I}_n)^{-1}\frac{\partial\mathcal{H}_{{\rm int},l}}{\partial\epsilon}\Big]\cdot\epsilon,
\end{eqnarray}
yielding $C_3=C_1+C_2$. This analysis holds for the case with more interaction terms, and shows that the term proportional to $\epsilon$ in the eigenequation of a system involving multiple interaction terms can be represented as the sum of such terms when considering each interaction term individually.


\end{document}